\documentclass[preprint,showpacs,preprintnumbers,amsmath,amssymb,superscriptaddress]{revtex4-1}
 \usepackage{graphicx}
 \usepackage{dcolumn}
 \usepackage{bm}
 \usepackage{ulem}
\usepackage{tensor}
\usepackage{axodraw}
\usepackage{xcolor}
\def\q{\bm{q}}
\def\p{\bm{p}}
\def\k{\bm{k}}
\usepackage{color}
\usepackage{ulem}  

\begin{document}

\title{Hadronic effects on the $cc\bar{q}\bar{q}$ tetraquark state in relativistic heavy
ion collisions}
\author{Juhee Hong}
\affiliation{Department of Physics and Institute of Physics and Applied Physics, Yonsei University,
Seoul 03722, Korea}
\author{Sungtae Cho}
\affiliation{Division of Science Education, Kangwon National
University, Chuncheon 24341, Korea}
\author{Taesoo Song}
\affiliation{Frankfurt Institute for Advanced Studies, Johann Wolfgang Goethe Universit\"{a}t, Frankfurt am Main, Germany}
\affiliation{Institut f\"{u}r Theoretische Physik, Universit\"{a}t
Gie\ss en, Germany}
\author{Su Houng Lee}
\affiliation{Department of Physics and Institute of Physics and Applied Physics, Yonsei University,
Seoul 03722, Korea}
\date{\today}

\begin{abstract}
We study the hadronic effects on the
$cc\bar{q}\bar{q}$ tetraquark state by focusing on the
$T_{cc}(1^+)$ meson during the hadronic stage of relativistic
heavy ion collisions. We evaluate the absorption cross section of
the $T_{cc}(1^+)$ meson by pions in the quasi-free approximation,
and investigate the time evolution of the $T_{cc}(1^+)$ abundance
in the hadronic medium based on the effective
volume and temperature of the hadronic phase at both RHIC and LHC
modelled by hydrodynamic calculations with the lattice equation of
state. We probe two possible scenarios for the structure of
$T_{cc}$, where it is assumed to be either a compact multiquark
state or a larger sized molecular configuration composed of
$DD^*$. Our numerical results suggest that the hadronic effects on
the $T_{cc}$ production is insignificant, and its final abundance
depends on the initial yield of $T_{cc}$ produced from the
quark-gluon plasma phase, which will depend on the assumed
structure of the state. 
\end{abstract}

\maketitle

\section{Introduction}

Exotic hadrons have been proposed to be important
probes in understanding the
fundamentals of the strong interaction in hadron physics
\cite{Jaffe:1976ig,Jaffe:1976ih}. The excitement in the subject
has restarted from the observation of $D_{sJ}(2317)$
\cite{exp1} and $X(3872)$ \cite{exp2}, whose masses did not fit
well within the conventional potential model approaches, and
continues to the present day with the recent observation of
$P_c(4380)^+$ and $P_c(4450)^+$ \cite{exp3}. Detailed theoretical
studies on the
structure and properties of these states have been reported using various models  
 \cite{review1,review2,review3,review4}. Moreover, it
has been argued that relativistic heavy ion
collisions provide an excellent venue to produce some of these and
previously proposed exotic states because they contain heavy
quarks, which are profusely produced in these experiments
\cite{exhic1, exhic2, exotic}.
Among many exotic hadrons, we focus here
on the proposed doubly charmed tetraquark
$T_{cc}(cc\bar{u}\bar{d}=DD^*)$ with the quantum number
$I(J^P)=0(1^+)$ \cite{potential,Lipkin:1986dw,Manohar:1992nd}.

There are several reasons why $T_{cc}$ is of particular interest.
First of all, it is a flavor exotic tetraquark, which has never
been observed before. Second, with the recent discovery of the
doubly charmed baryon at CERN \cite{Aaij:2017ueg}, the possibility
of observing a similar doubly charmed hadron with the light quark
replaced by a strongly correlated light anti-diquark seems quite
plausible. Finally, analyzing the structure of this particle in
the constituent quark model, one finds that this particle is the
only candidate where there is a strong attraction in the compact
configuration compared to two separated meson. This is so because
while previously observed exotic candidates such as the $X(3872)$
is composed of $q\bar{q} Q\bar{Q}$, where $q,Q$ are light and
heavy quarks respectively, the proposed $T_{cc}$ state is composed
of $QQ \bar{q} \bar{q}$ quarks. The latter quark structure favors
a compact tetraquark configuration as the additional light anti-diquark structure
$\bar{q} \bar{q}$ in the isospin zero channel provides an
attraction larger than that for  the two $Q \bar{q}$ in a
separated meson configurations
\cite{Park:2013fda,Hyodo:2017hue,Luo:2017eub}. Hence, $T_{cc}$ is
a unique multiquark candidate state that could be compact.

The measured yields of ground state particles and their ratios
from relativistic heavy ion collisions can be well described by
statistical models \cite{stat1,stat,Stachel:2013zma}. On the other
hand, there are indications that yields for resonances with
structures different from ground states, deviate from the
statistical model prediction
\cite{Kanada-Enyo:2006dxd,Cho:2014xha}. In particular, it was
argued that the yields of compact multiquark configurations would
be an order of magnitude suppressed compared to a molecular
configuration or a usual hadron with the same quantum number and
mass, if allowed, which should follow the statistical model
prediction \cite{exhic1,exhic2,exotic}. However, these results
were obtained without considering the hadronic effects, which
could change the initial production rate at the chemical
freeze-out due to the interaction with other particles during the
hadronic expansion before the kinetic freeze-out. The importance
of this effect has been confirmed for states with large intrinsic
width such as the $K^*$, which has been observed both at RHIC and
LHC with yield ratios to the $K$ that are systematically
reduced compared to the statistical model predictions
\cite{Adam:2017zbf}. If the hadronic effects are large, the hope
of using production yields to discriminate the structure of an
exotic particle through its production could be problematic. In
fact, for similar reasons, the hadronic effects of exotic
candidates have been estimated for the $D_{sJ}(2317)$ \cite{ko}
and $X(3872)$ \cite{cho,pion}.

In this work, we estimate the hadronic effects on the
$T_{cc}$ yields in heavy ion collisions to assess if the initial
yields at the hadronization point is maintained so that its
structure can be discriminated. We also solve a
hydrodynamic model based on the lattice equation of state with and
without viscosity, and parameterize the resulting time 
dependence of the temperature and volume during the hadronic phase
at both RHIC and LHC that will be used in this and in similar
future works.

This work is organized as follows. In Section \ref{hydro}, we
introduce a simplified hydrodynamic model to calculate and
parameterize the time dependence of the temperature and volume of the
hadronic phase at RHIC and LHC. In Section \ref{hadronization}, we
discuss the hadronization in relativistic heavy ion collisions and
the $T_{cc}$ yields calculated in two possible scenarios, where
$T_{cc}$ is either a compact configuration with suppressed yield
estimated within the coalescence model or a weakly bound molecular
configuration that should follow the statistical model prediction.
In Section \ref{cross_section}, the cross sections of the $T_{cc}$
absorption by pions are calculated in the quasifree approximation.
In Section \ref{evolution}, the time evolution of the $T_{cc}$
abundance is studied by solving the rate equation in the two
possible scenarios. In Section \ref{finalstates}, we give possible
production final states that can be used to observe these states
from heavy ion collisions. Finally, we summarize our results in
Section \ref{summary}.

\section{Hydrodynamic equation for the hadronic phase}
\label{hydro}

Hydrodynamic equations are given by $\partial_\mu
T^{\mu\nu}=0$, where the energy-momentum tensor
$T^{\mu\nu}=(e+p)u^\mu u^\nu-pg^{\mu\nu}+\pi^{\mu\nu}$ with $e$,
$p$, $u^\mu$, and $\pi^{\mu\nu}$ being, respectively, the energy
density, pressure, four-velocity of flow, and the traceless symmetric
shear tensor. For simplicity, we assume the boost-invariance and
consider central collisions, that is, symmetric expansion in the
transverse plane. Then there are only two independent hydrodynamic
equations \cite{Heinz:2005bw}:
\begin{eqnarray}
\frac{1}{\tau}\partial_\tau(\tau T^{\tau \tau})+\frac{1}{r}
\partial_r(r T^{r \tau})&=&-\frac{1}{\tau}(p+\tau^2\pi^{\eta\eta})\, ,
\label{energy6}\\
\frac{1}{\tau}\partial_\tau(\tau T^{\tau r})+\frac{1}{r}
\partial_r(r T^{r r})&=&\frac{1}{r}(p+r^2\pi^{\phi\phi}) \, ,
\label{momentum}
\end{eqnarray}
in the $(\tau, r, \phi, \eta)$ coordinate system defined by
\begin{eqnarray}
\tau&=&\sqrt{t^2-z^2} \, , ~~~~~\eta=\frac{1}{2}\ln \frac{t+z}{t-z}
\, ,\nonumber\\
r&=&\sqrt{x^2+y^2} \, , ~~~~\phi=\tan^{-1}(y/x) \, .
\end{eqnarray}
Nonvanishing energy-momentum tensors and shear tensors are
respectively expressed as \cite{Heinz:2005bw}
\begin{eqnarray}
T^{\tau\tau}&=&(e+P_r)u_\tau^2 -P_r \, ,\nonumber\\
T^{\tau r}&=&(e+P_r)u_\tau u_r \, ,\nonumber\\
T^{r r}&=&(e+P_r)u_r^2+P_r \, ,
\end{eqnarray}
where $P_r\equiv p-\tau^2\pi^{\eta\eta}-r^2\pi^{\phi\phi}$ is the
effective radial pressure, and
\begin{eqnarray}
\pi^{\tau r}&=&v_r\pi^{rr} \, ,\nonumber\\
\pi^{\tau\tau}&=&v_r\pi^{\tau r}=v_r^2\pi^{rr} \, ,\nonumber\\
\pi^{rr}&=&-\gamma_r^2(r^2\pi^{\phi\phi}+\tau^2\pi^{\eta\eta}) \, ,
\end{eqnarray}
with $v_r$ being the radial velocity and the shear tensors $\pi^{\phi\phi}$ and $\pi^{\eta\eta}$ being
the only independent ones. The components 
$\pi^{\phi\phi}$ and $\pi^{\eta\eta}$ are boost-invariant in the
radial direction and satisfy the following simplified
Israel-Stewart equations:
\begin{eqnarray}
(\partial_\tau +v_r \partial_r)\pi^{\eta \eta}&=&-\frac{1}{\gamma_r
\tau_\pi}\bigg[\pi^{\eta \eta}-\frac{2\eta_s}{\tau^2}\bigg(\frac{
\theta}{3}-\frac{\gamma_r}{\tau}\bigg)\bigg] \, ,\label{shear1a}\\
(\partial_\tau +v_r \partial_r)\pi^{\phi \phi}
&=&-\frac{1}{\gamma_r \tau_\pi}\bigg[\pi^{\phi \phi}-
\frac{2\eta_s}{r^2}\bigg(\frac{\theta}{3}-\frac{\gamma_r
v_r}{r}\bigg)\bigg] \, , \label{shear1b}
\end{eqnarray}
where
\begin{eqnarray}
\theta=\partial\cdot u=\frac{1}{\tau}\partial_\tau (\tau
\gamma_r)+ \frac{1}{r}\partial_r(rv_r \gamma_r) \, ,\nonumber
\end{eqnarray}
with $\eta_s$ and $\tau_\pi$ being the shear viscosity and the
relaxation time for the particle distributions, respectively.
Furthermore, the condition $u_\mu (T_{;\nu}^{\nu \mu})=0$, where
$T_{;\nu}^{\nu \mu}$ is the covariant derivative and the flow
velocity $(u_\tau, u_r, u_\phi, u_\eta)=(\gamma/\cosh\eta,\gamma
v_r,0,0)$ reduces to $(\gamma_r,\gamma_r v_r,0,0)$ with
$\gamma_r=1/\sqrt{1-v_r^2}$ in midrapidity, leads to
\begin{eqnarray}
&&\frac{1}{\tau}\partial_\tau (\tau s \gamma_r)+\frac{1}{r}\partial_r
(rs\gamma_r v_r)=-\frac{1}{T}\bigg[\frac{u_\tau}{\tau}\tau^2\pi^{\eta\eta}
\nonumber\\
&&\qquad\qquad\qquad\qquad\qquad
+\frac{u_r}{r}r^2\pi^{\phi\phi}-(\partial_\tau u_\tau+\partial_r
u_r) (r^2\pi^{\phi\phi}+\tau^2\pi^{\eta \eta})\bigg] \, ,
\label{entropy6}
\end{eqnarray}
where $s=(e+p)/T$ is the local entropy density in the hot dense
matter. Eq.~(\ref{entropy6}) shows that the total entropy is not
conserved in the presence of nonzero shear tensors.

Integrating Eqs. (\ref{energy6}), (\ref{shear1a}),
(\ref{shear1b}), and (\ref{entropy6}) over the transverse plane,
we have \cite{Song:2010fk}
\begin{eqnarray}
&&\partial_\tau (A\tau \langle T^{\tau \tau}\rangle)=-(p+\pi^\eta_\eta)A
\, ,\label{energy7}\\
\nonumber\\
&&\frac{T}{\tau}\partial_\tau (A\tau s \langle \gamma_r\rangle)=
-A\bigg\langle\frac{\gamma_r v_r}{r}\bigg\rangle \pi^\phi_\phi-
\frac{A\langle \gamma_r\rangle}{\tau}\pi^\eta_\eta\nonumber\\
&&~~~~~~~\qquad\qquad\qquad\qquad\qquad
+\bigg[\partial_\tau(A\langle \gamma_r\rangle)-\frac{\gamma_R
\dot{R}}{R}A\bigg](\pi^\phi_\phi+\pi^\eta_\eta) \, ,\label{entropy7}\\
\nonumber\\
&&\partial_\tau (A\langle \gamma_r\rangle \pi^\eta_\eta) -\bigg[
\partial_\tau(A\langle\gamma_r\rangle)+2\frac{A\langle\gamma_r\rangle}
{\tau} \bigg]\pi^\eta_\eta\nonumber\\
&&~~~~\qquad\qquad\qquad\qquad\qquad
=-\frac{A}{\tau_\pi}\bigg[\pi^\eta_\eta-2\eta_s\bigg(\frac{
\langle\theta\rangle}{3}-\frac{\langle\gamma_r\rangle}{\tau}\bigg)
\bigg] \, ,\label{entropy7b}\\
\nonumber\\
&&\partial_\tau(A\langle\gamma_r\rangle~ \pi^\phi_\phi)-\bigg[\partial_\tau(A\langle\gamma_r\rangle)+2A\bigg\langle\frac{\gamma_r v_r}
{r}\bigg\rangle\bigg]\pi^\phi_\phi\nonumber\\
&&~~~~\qquad\qquad\qquad\qquad\qquad
=-\frac{A}{\tau_\pi}\bigg[ \pi^\phi_\phi-2\eta_s \bigg(\frac{
\langle\theta\rangle}{3}-\bigg\langle\frac{\gamma_r v_r}{r}\bigg
\rangle\bigg)\bigg]\label{shear7b} \, ,
\end{eqnarray}
where $A=\pi R^2$ ($R$ is the radius of nuclear matter), 
$\langle T^{\tau\tau}\rangle=\int dA \,
T^{\tau\tau}/A=(e+p)\langle \gamma_r^2\rangle-p, \, \langle u^\tau
\rangle=\langle \gamma_r \rangle$, 
$\pi^\eta_\eta\equiv\tau^2\pi^{\eta\eta}$, and $\pi^\phi_\phi\equiv
r^2\pi^{\phi\phi}$. We note that the total derivatives with
respect to $r$ disappear due to the boundary condition. Assuming
that the radial flow velocity is a linear function of the radial
distance from the center, that is, $\gamma_r v_r=\gamma_R
\dot{R}(r/R)$, where $\dot{R}=\partial R/\partial \tau$ and
$\gamma_R=1/\sqrt{1-\dot{R}^2}$,
\begin{eqnarray}
\langle\gamma_r^2\rangle&=&1+\frac{\gamma_R^2 \dot{R}^2}{2}
\, ,\nonumber\\
\langle\gamma_r^2 v_r^2\rangle&=&\frac{\gamma_R^2 \dot{R}^2}{2}
\, ,\nonumber\\
\langle\gamma_r\rangle&=&\frac{2}{3\gamma_R^2 \dot{R}^2}
\left(\gamma_R^3-1\right) \, ,\nonumber\\
\bigg\langle\frac{\gamma_r v_r}{r}\bigg\rangle&=&\frac{\gamma_R
\dot{R}}{R} \, .
\label{gamma}
\end{eqnarray}
Here, we make the assumption that nuclear matter has a definite
boundary and $e$, $s$, and $p$ are uniform inside. In real
hydrodynamic simulations, the energy-momentum tensor is numerically
calculated for all time-space cells leading to a different
temperature for each cell so that the hypersurface for a constant
temperature has a complex structure in the $xy\tau$-space.
But at the same time, one finds that most of the points composing
the hypersurface are located on a semiconstant $\tau$ plane
\cite{Song:2011kw}. That is why the blast wave model had been
successful and widely used before sophisticated hydrodynamics
became popular. This is the basis for our approximation.

We then numerically solve simultaneous Eqs. (\ref{energy7}) to
(\ref{shear7b}) by using the lattice equation of state
\cite{Song:2010fk, Borsanyi:2010cj}. The ratio of the shear
viscosity to entropy density is taken to be $1/(4\pi)$ for QGP
\cite{Kovtun:2004de}, and ten times this value for hadron gas
\cite{Demir:2008tr}. For the relaxation time $\tau_\pi$, we assume
$\eta/\tau_\pi=sT/3$ for both QGP and hadron
gas \cite{Song:2009gc}. The initial thermalization time for
hydrodynamic simulations is assumed to be 0.5 fm/c, and the
initial radius is given by the transverse area where the local
temperature is above 150 MeV. Although the hydrodynamic approach is
marginal in the hadron gas phase, it has successfully reproduced
abundant experimental data from relativistic heavy ion collisions
\cite{Kolb:2000sd,Schenke:2010nt}.

\begin{figure}
\includegraphics[width=0.45\textwidth]{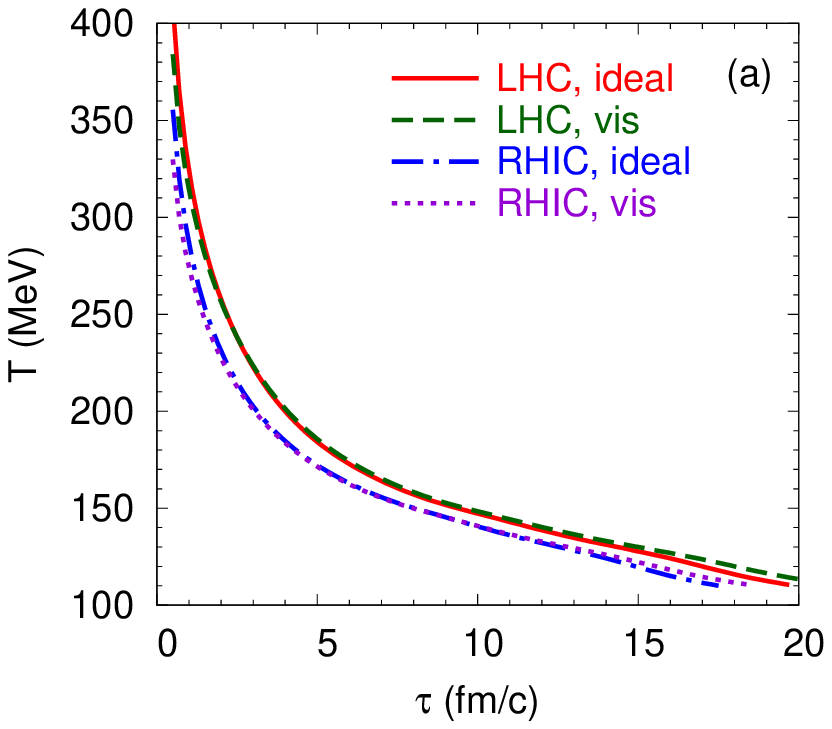}
\includegraphics[width=0.45\textwidth]{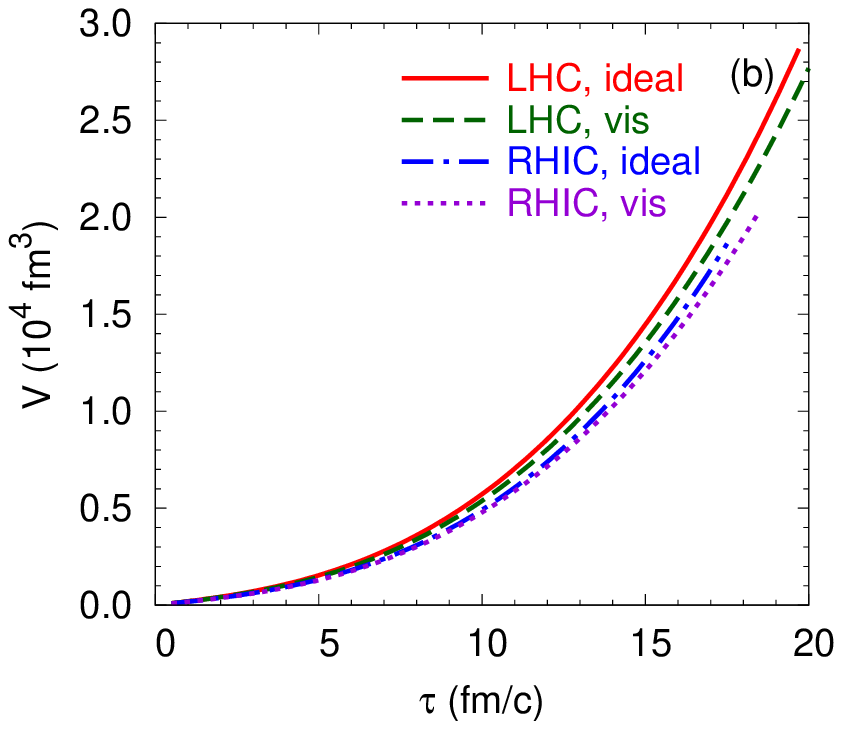}
\caption{(a) Temperature and (b) volume for LHC and RHIC during
the hadronic expansion \cite{exotic}. } \label{TVfig}
\end{figure}

According to the hydrodynamic calculations, the temperature and volume 
during the hadronic phase for LHC and RHIC change with time as shown 
in Fig. \ref{TVfig}. 
We now parameterize the results for the $\tau$ dependence of the
volume and temperature using the following form \cite{TVmodel,ko}:
\begin{eqnarray}
\label{TVeq}
V(\tau)&=&\pi\left[R+v(\tau-\tau_C)+\frac{a}{2}(\tau-\tau_C)^2\right]^2
c\tau \, , \nonumber\\
T(\tau)&=&T_C-(T_H-T_F)\left(\frac{\tau-\tau_H}{\tau_F-\tau_H}\right)^\alpha
\qquad\mbox{for $\tau>\tau_H$} \, ,
\end{eqnarray}
with $T_c \, (\tau_c)$, $T_H \, (\tau_H)$, and $T_F\, (\tau_F)$ being
the critical, hadronization, and kinetic freeze-out temperature
(time), respectively. In Eq. (\ref{TVeq}), we take
$T_H=156\, (162)$ MeV, $T_F=115 \, (119)$ MeV for LHC (RHIC), and 
$T_C=T_H$ by following the first scenario of
Ref. \cite{exotic}. $R$, $v$, $a$, and $\alpha$ 
have been treated as fitting parameters. All the parameters
used in the model are given in Table \ref{TVcoeff}. 

\begin{table}
\centering
\caption{
Parameters used in the phenomenological model of Eq. (\ref{TVeq}).
}
\newcolumntype{C}{>{\centering\arraybackslash}p{9.1ex}}
\begin{tabular}{ C C C  C C C C C C C}
\hline
\hline
& & $T_C=T_H$  & $T_F$ & $\tau_C=\tau_H$ & $\tau_F$ & $R$ & $v$ & $a$ & $\alpha$\\
 &  & (MeV)  & (MeV)       &  (fm/c) & (fm/c) & (fm)& (c)& ($c^2$/fm) &\\
\hline
LHC &ideal & 156  & 115 & 8.1 & 18.3 & 12.1 & 0.70 & 0.022 & 0.95 \\
 &viscous & 156  & 115 & 8.3 & 19.5 & 11.9 & 0.67 & 0.020 & 0.93 \\
RHIC & ideal&162  & 119 & 6.1 & 15.1 & 9.9 & 0.59 & 0.030 & 0.85 \\
 & viscous &162  & 119 & 6.1 & 15.7 & 9.8 & 0.58 & 0.024 & 0.79 \\
\hline
\hline
\end{tabular}
\label{TVcoeff}
\end{table}

\section{Hadronization in relativistic heavy ion collisions}
\label{hadronization}

We assume that conventional hadrons such as $\pi$, $D$ and $D^*$
are in chemical and thermal equilibrium when they are produced at
the chemical freeze-out. 
The abundance of a particle in equilibrium is statistically given by \cite{statrev}
\begin{eqnarray}
\label{stateq}
N_i^{eq}(\tau)&=&g_i\gamma_iV(\tau)\int
\frac{d^3\p}{(2\pi)^3}f(\p) \, ,
\nonumber\\
&=& \frac{1}{2\pi^2} \, g_i \gamma_i  m_i^2 \, V(\tau) \,
T(\tau) \, K_2\left(\frac{m_i}{T(\tau)}\right) \, ,
\end{eqnarray}
where $g_i=(2S_i+1)(2I_i+1)$ is the spin and isospin degeneracy
and $\gamma_i$ is the fugacity. In the second line, the Boltzmann
distribution $f(\p)=\exp[-\sqrt{\p^2+m_i^2}/T(\tau)]$ has
been used and $K_2$ is the modified Bessel function of the second
kind. For simplicity we ignore the correction term to $f(\p)$ for 
shear viscosity. Since the production and annihilation cross
sections of charm quarks are small \cite{lag,charm1,charm2}, the
number of charm quarks is conserved during the time evolution
of the hadronic matter. 
From the total number of charm quarks, $N_c=11 \, (4.1)$
\cite{exotic}, the charm fugacity is determined as
$\gamma_c=51 \, (22)$ for LHC (RHIC). Here the charm fugacity is
slightly different from that in Ref.~\cite{exotic} because we use
only $D,D^*,D_s,\, \mbox{and}\, D^*_s$ to saturate the charm quarks as in
Eq.~(\ref{charm-fugacity}).  By following Refs. \cite{ko,pion}, the
number of pions at RHIC is set to be $926$ at the kinetic
freeze-out. For that purpose, we introduce a pion chemical
potential with effective fugacity of 1.4 and use the same factor
at LHC. This effect is to include the feed-down contributions from
excited states such as the omega, delta, and $K^*$. Although these
pions will only have a limited contribution to the absorption
during hadronic phase, we will include them in the calculation to
allow for the maximum effect. 

If the $T_{cc}$ is of a molecular configuration composed of
a weakly bound $D D^*$, the production yield is
expected to follow the statistical model 
prediction as the production yield of light
nuclei do so. In such a case, the
number of the doubly charmed $T_{cc}$ is given by Eq.
(\ref{stateq}) with $\gamma_c^2$, $V(\tau_H)$ and $T(\tau_H)$ for the
fugacity,  volume and temperature, respectively. On the other
hand, if the $T_{cc}$ is a compact multiquark state with
the size of a usual hadron, then the production 
yield would be suppressed compared to the statistical model
prediction. The production yields have been estimated by the
coalescence model, whose parameters have been fitted to reproduce
the ground state hadron yields \cite{exotic}. The two cases are
summarized in Table \ref{NTcc}. Throughout this paper, we use the
average masses: $m_\pi=137.3$ MeV, $m_D=1867.2$ MeV,
and $m_{D^*}=2008.6$ MeV \cite{pdg}.

\begin{table}
\centering \caption{$T_{cc}$ yields at hadronization.}
\newcolumntype{C}{>{\centering\arraybackslash}p{22ex}}
\begin{tabular}{ C  C  C}
\hline
\hline
 & molecular  & compact multiquark  \\
\hline
LHC & $2.0\times10^{-3}$  & $1.1\times10^{-4}$   \\
RHIC & $5.1\times10^{-4}$  & $5.0\times10^{-5}$   \\
\hline
\hline
\end{tabular}
\label{NTcc}
\end{table}

\section{$T_{cc}$ absorption cross sections}
\label{cross_section}

The $T_{cc}$ can be produced or destroyed by interacting
with other comoving particles during the hadronic expansion stage.
Since pions are the most abundant particles with small mass, the
interaction with them is the main contribution to the $T_{cc}$
abundance. In this section, we calculate the absorption cross
sections of the $T_{cc}$ by pions in the quasifree
approximation.

\begin{figure}
\begin{center} \begin{picture}(350,90)(0,70)
\Text(21,130)[r]{\small(a)}
\Line(-20,120)(5,90)
\Line(25,90)(50,120)
\Line(-20,90)(50,90)
\Line(-20,70)(50,70)
\Text(-20,125)[r]{\small$\pi$}
\Text(58,125)[r]{\small$\pi$}
\Text(23,82)[r]{\small$D^*$}
\Text(-22,90)[r]{\small$D$}
\Text(-20,70)[r]{\small$D^*$}
\Text(60,90)[r]{\small$D$}
\Text(65,70)[r]{\small$D^*$}
\Text(131,130)[r]{\small(b)}
\Line(95,120)(140,90)
\Line(110,90)(155,120)
\Line(90,90)(160,90)
\Line(90,70)(160,70)
\Text(95,125)[r]{\small$\pi$}
\Text(163,125)[r]{\small$\pi$}
\Text(133,82)[r]{\small$D^*$}
\Text(88,90)[r]{\small$D$}
\Text(90,70)[r]{\small$D^*$}
\Text(171,90)[r]{\small$D$}
\Text(176,70)[r]{\small$D^*$}
\Text(241,130)[r]{\small(c)}
\Line(200,120)(225,90)
\Line(245,90)(270,120)
\Line(200,90)(270,90)
\Line(200,70)(270,70)
\Text(200,125)[r]{\small$\pi$}
\Text(278,125)[r]{\small$\pi$}
\Text(240,82)[r]{\small$D$}
\Text(200,90)[r]{\small$D^*$}
\Text(200,70)[r]{\small$D$}
\Text(285,90)[r]{\small$D^*$}
\Text(281,70)[r]{\small$D$}
\Text(351,130)[r]{\small(d)}
\Text(315,125)[r]{\small$\pi$}
\Text(383,125)[r]{\small$\pi$}
\Text(350,82)[r]{\small$D$}
\Text(312,90)[r]{\small$D^*$}
\Text(310,70)[r]{\small$D$}
\Text(396,90)[r]{\small$D^*$}
\Text(392,70)[r]{\small$D$}
\Line(315,120)(360,90)
\Line(330,90)(375,120)
\Line(310,90)(380,90)
\Line(310,70)(380,70)
\end{picture}
\end{center}
\caption{Diagrams contributing to the $T_{cc}$ abundance. In the
quasifree approximation, (a) and (b) correspond to the elastic
scattering $D+\pi\rightarrow D+\pi$, and (c) and (d) to
$D^*+\pi\rightarrow D^*+\pi$. } \label{diagram}
\end{figure}
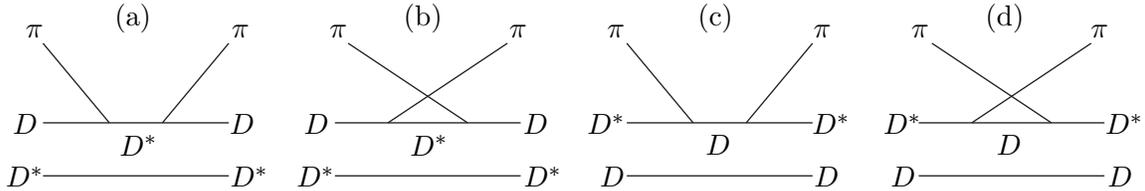

The quasifree approximation has been previously used to estimate
the dissociation of charmonia by partons \cite{ralf}. The
approximation was shown to be valid when the binding energies of
charmonia are small at high temperature, and $c$ and $\bar{c}$
quarks inside charmonia can be treated like quasifree particles
\cite{quasifree} (see Appendix \ref{appendix-23} for the details).
In fact, for the charmonium case, it can be seen that an exact
next to leading order QCD calculation allowing for the compact
size gives a similar result for the thermal width
\cite{Park:2007zza} as that obtained using the quasifree
approximation when the process involves the same number of initial
and final states. Here, we estimate the dissociation
cross section of the $T_{cc}$ by pions by estimating the
$D$ and $D^*$ components of the $T_{cc}$ in two possible
scenarios under the quasifree
approximation.

In the quasifree approximation, the cross section of
$T_{cc}+\pi\rightarrow D+D^*+\pi$ can be evaluated by adding the
elastic scattering $D+\pi\rightarrow D+\pi$ and $D^*+\pi\rightarrow
D^*+\pi$ (see Fig. \ref{diagram}). For the effective interaction
vertices, we use the following interaction Lagrangian \cite{lag}:
\begin{equation}
\label{Leff} \mathcal{L}_{\pi DD^*}=ig_{\pi DD^*}D^{*\mu}
\bm{\tau}\cdot(\bar{D} \partial_\mu
\bm{\pi}-\partial_\mu\bar{D}\bm{\pi})+\mbox{h.c.} \, ,
\end{equation}
where $\bm{\tau}$ are the Pauli matrices, $\bm{\pi}$ is the pion
isospin triplet, and $D=(D^0, D^+)$ and $D^*=(D^{*0},D^{*+})$ are
the pseudoscalar and vector charm meson doublets, respectively.
The meson coupling $g_{\pi DD^*}$ is determined from the
$D^*\rightarrow D\pi$ decay width
\begin{equation}
\Gamma_{D^*\rightarrow D\pi}=
\frac{g_{\pi DD^*}^2 p_{cm}^3}{2\pi m_{D^*}^2} \, ,
\end{equation}
where $p_{cm}$ is the momentum in the center of mass frame. 
By comparing with the experimental data, the full width
$\Gamma_{D^*\rightarrow D\pi}=83.4$ keV \cite{pdg}, we obtain
$g_{\pi DD^*}\simeq7.8$.


The scattering amplitude of the process
$D(p_1)+\pi(p_2)\rightarrow D(p_3)+\pi(p_4)$ is then given as
\begin{equation}
\mathcal{M}_{D\pi\rightarrow D\pi}= \mathcal{M}^{(a)}+\mathcal{M}^{(b)} \, ,
\end{equation}
where
\begin{eqnarray}
\label{mtx1}
\mathcal{M}^{(a)}&=&\frac{2^{N(\pi^\pm)/2}g_{\pi DD^*}^2}{s-m_{D^*}^2}
\left[-g^{\mu\nu}+\frac{(p_1+p_2)^\mu(p_1+p_2)^\nu}{m_{D^*}^2}\right]
(p_1-p_2)_\mu (p_3-p_4)_\nu \, ,
\nonumber\\
\mathcal{M}^{(b)}&=&\frac{2^{N(\pi^\pm)/2}g_{\pi DD^*}^2}{u-m_{D^*}^2}
\left[-g^{\mu\nu}+\frac{(p_1-p_4)^\mu(p_1-p_4)^\nu}{m_{D^*}^2}\right]
(p_1+p_4)_\mu (p_2+p_3)_\nu \, .
\end{eqnarray}
Here, $N(\pi^\pm)$ is the number of charged pions involved in initial and 
final states of the process (see Table \ref{process}).

For $D^*(p_1)+\pi(p_2)\rightarrow D^*(p_3)+\pi(p_4)$, we have
\begin{equation}
\mathcal{M}_{D^*\pi\rightarrow D^*\pi}= \mathcal{M}^{(c)}+\mathcal{M}^{(d)} \, ,
\end{equation}
with
\begin{eqnarray}
\label{mtx2}
\mathcal{M}^{(c)}&=&-\frac{2^{N(\pi^\pm)/2}g_{\pi DD^*}^2
\epsilon_1^\mu\epsilon_3^{*\nu}}
{s-m_{D}^2}(p_1+2p_2)_\mu(p_3+2p_4)_\nu \, ,
\nonumber\\
\mathcal{M}^{(d)}&=&-\frac{2^{N(\pi^\pm)/2}g_{\pi DD^*}^2
\epsilon_1^\mu\epsilon_3^{*\nu}}
{u-m_{D}^2}(-p_1+2p_4)_\mu(2p_2-p_3)_\nu \, .
\end{eqnarray}

In the center of mass frame, the spin and isospin averaged cross
section is
\begin{equation}
\label{sigma}
\sigma=\frac{1}{64\pi^2g_1g_2s}\frac{|\p_f|}{|\p_i|}\int d\Omega
\sum_{S,I}|\mathcal{M}|^2F^4 \, ,
\end{equation}
where $g_1$ and $g_2$ are the degeneracies of initial particles,
$\p_i$ ($\p_f$) is the spatial momentum of initial (final)
particles, and the summation is over the spins and isospins of
both initial and final particles. The relevant processes are
listed in Table \ref{process}. At each interaction vertex, we have
used the following form factors:
\begin{equation}
F=\frac{\Lambda^2}{\Lambda^2+(\omega^2-m_{ex}^2)}
\qquad \mbox{and}
\qquad \frac{\Lambda^2}{\Lambda^2+\q^2} \, ,
\end{equation}
for the $s$- and $u$-channels, respectively. Here, the cutoff
$\Lambda=1.0$ GeV is used, $m_{ex}$ is the mass of the exchanged particle, 
$\omega$ is the total energy of
incoming particles in the s-channel, and $\q$ is the momentum
transfer in the u-channel in the center of mass frame. Using the
form factors, the cross sections do not increase with the total
center of mass energy.

\begin{table}
\centering \caption{$2\rightarrow 2$ processes contributing to the
spin and isospin averaged cross section of Eq. (\ref{sigma}). With
the effective Lagrangian of Eq. (\ref{Leff}), the matrix elements
involve the factor $2^{N(\pi^\pm)/2}$ in Eqs. (\ref{mtx1})
and (\ref{mtx2}). }
\newcolumntype{C}{>{\centering\arraybackslash}p{20ex}}
\begin{tabular}{ C  C  C C }
\hline
\hline
process &  diagram  & process & diagram \\
\hline
 $D^+\pi^0\rightarrow D^+\pi^0$
& (a)+(b)
& $D^{*+}\pi^0\rightarrow D^{*+}\pi^0$
& (c)+(d)
 \\
 $D^+\pi^0\rightarrow D^0\pi^+$
& (a)+(b)
& $D^{*+}\pi^0\rightarrow D^{*0}\pi^+$
& (c)+(d)
 \\
 $D^+\pi^-\rightarrow D^+\pi^-$
& (a)
& $D^{*+}\pi^-\rightarrow D^{*+}\pi^-$
& (c)
 \\
 $D^+\pi^-\rightarrow D^0\pi^0$
& (a)+(b)
& $D^{*+}\pi^-\rightarrow D^{*0}\pi^0$
& (c)+(d)
 \\
 $D^+\pi^+\rightarrow D^+\pi^+$
& (b)
& $D^{*+}\pi^+\rightarrow D^{*+}\pi^+$
& (d)
 \\
 $D^0\pi^+\rightarrow D^0\pi^+$
& (a)
& $D^{*0}\pi^+\rightarrow D^{*0}\pi^+$
& (c)
 \\
 $D^0\pi^+\rightarrow D^+\pi^0$
& (a)+(b)
& $D^{*0}\pi^+\rightarrow D^{*+}\pi^0$
& (c)+(d)
 \\
 $D^0\pi^0\rightarrow D^0\pi^0$
& (a)+(b)
& $D^{*0}\pi^0\rightarrow D^{*0}\pi^0$
& (c)+(d)
 \\
 $D^0\pi^0\rightarrow D^+\pi^-$
& (a)+(b)
& $D^{*0}\pi^0\rightarrow D^{*+}\pi^-$
& (c)+(d)
 \\
 $D^0\pi^-\rightarrow D^0\pi^-$
& (b)
& $D^{*0}\pi^-\rightarrow D^{*0}\pi^-$
& (d)
 \\
\hline
\hline
\end{tabular}
\label{process}
\end{table}

To take into account the thermal effects, we define 
$\langle \sigma_{ab\rightarrow cd}v_{ab}\rangle$, the
product of the cross section of two-body scattering
($ab\rightarrow cd$) and the relative velocity between initial
particles, 
$v_{ab}=\sqrt{(p_a\cdot p_b)^2-m_a^2m_b^2}/(E_aE_b)$, 
averaged over the thermal momentum distributions of
initial particles \cite{sv1,sv2}:
\begin{eqnarray}
\langle \sigma_{ab\rightarrow cd}v_{ab}\rangle(\tau)
&=&\frac{\int d^3\p_ad^3\p_b \, f_a(\p_a)f_b(\p_b) \, \sigma_{ab\rightarrow cd}v_{ab}}
{\int d^3\p_a d^3\p_b \, f_a(\p_a)f_b(\p_b)} \, ,
\nonumber\\
&=&\left[4\left(\frac{m_a}{T(\tau)}\right)^2\left(\frac{m_b}{T(\tau)}\right)^2 \,
K_2\left(\frac{m_a}{T(\tau)}\right)
K_2\left(\frac{m_b}{T(\tau)}\right)
\right]^{-1}
\int_{z_0} dz\, \sigma(\sqrt{s}=zT(\tau))
\nonumber\\
&&\qquad \times
\left[z^2-\left(\frac{m_a+m_b}{T(\tau)}\right)^2\right]
\left[z^2-\left(\frac{m_a-m_b}{T(\tau)}\right)^2\right]
K_1(z) \, ,
\label{thermal-average}
\end{eqnarray}
where $z_0=\mbox{max}[(m_a+m_b)/T(\tau),(m_c+m_d)/T(\tau)]$. It
should be noted, however, that we are approximating
$\sigma_{T_{cc} \pi \rightarrow DD^* \pi}$ by $\sigma_{D \pi
\rightarrow D \pi}$ and $\sigma_{D^* \pi \rightarrow D^* \pi}$.
Hence, when taking the thermal distribution, the distribution
$f_a( \p_a)$ should be that of the $T_{cc}$. Furthermore, the
threshold should also involve that of $m_D+m_{D^*} \rightarrow
m_{T_{cc}}$. This amounts to taking $m_a=m_c= m_{T_{cc}}$ instead
of $m_D$ or $m_{D^*}$. The same approximation will be taken when
calculating the inverse process. The derivation of this
formula is given in Appendix \ref{appendix-rate}.

\begin{figure}
\includegraphics[width=0.45\textwidth]{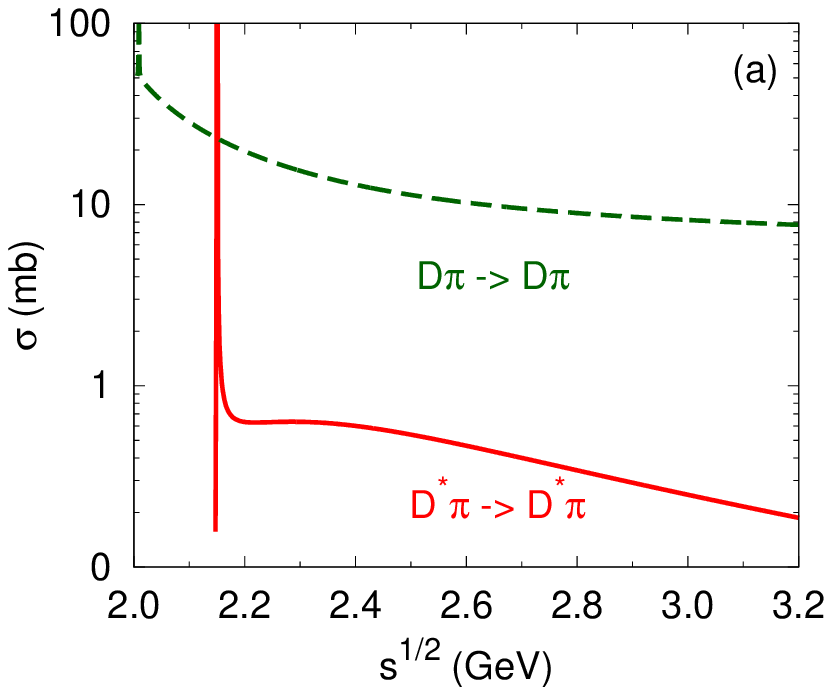}
\includegraphics[width=0.45\textwidth]{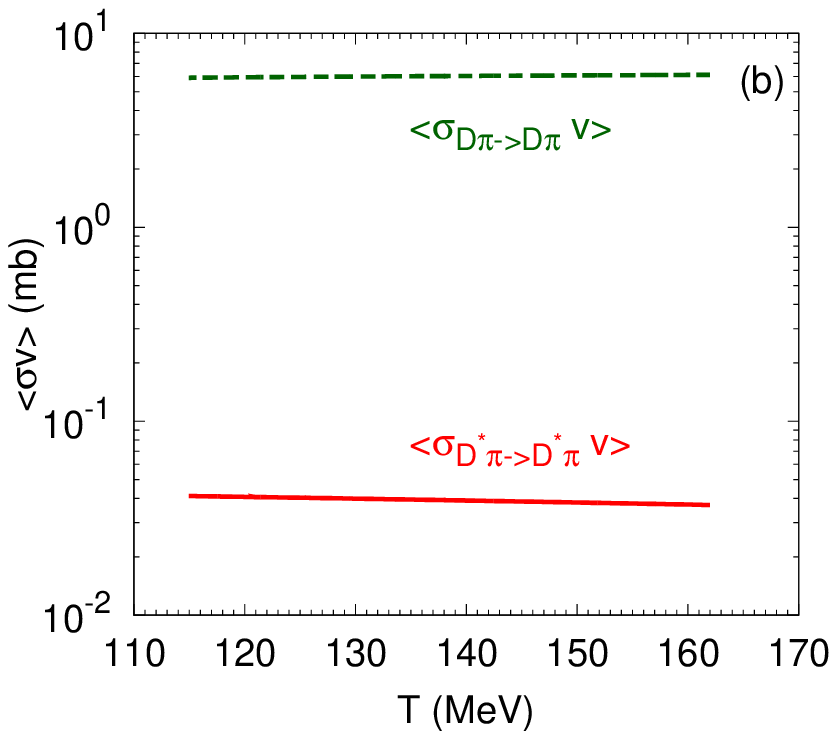}
\caption{ (a) The cross sections as functions of the total center
of mass energy and (b) the thermally averaged cross sections
contributing to the absorption of the $T_{cc}$ by pions. }
\label{cross}
\end{figure}
Fig. \ref{cross} shows the cross sections and the thermally
averaged ones of the elastic scattering $D(D^*)+\pi\rightarrow
D(D^*)+\pi$. The cross section of the s-channel in the process
$D+\pi\rightarrow D+\pi$ has a peak near the threshold energy
$\sqrt{s}_0=m_D+m_\pi$ since $m_D+m_\pi\approx m_{D^*}$.
Similarly, the cross section of the u-channel in
$D^*+\pi\rightarrow D^*+\pi$ diverges near
$\sqrt{s}_0=m_{D^*}+m_\pi$.

\section{Time evolution of the $T_{cc}$ abundance}
\label{evolution}

We consider the time evolution of the $T_{cc}$
abundance governed by (see Appendix \ref{appendix-rate})
\begin{multline}
\label{evolve}
\frac{dN_{T_{cc}}(\tau)}{d\tau}=
\langle \sigma_{T_{cc}\pi\rightarrow DD^*\pi}
v_{T_{cc}\pi}\rangle(\tau) \, n_\pi(\tau) \bigg[- N_{T_{cc}}(\tau) +N^{eq}_{T_{cc}}(\tau)
\,  \frac{N_D(\tau) \, N_{D^*}(\tau)}{N^{eq}_D(\tau) \, N^{eq}_{D^*}(\tau)} \, \bigg] ,
\end{multline}
where $n_\pi(\tau)=N_\pi(\tau)/V(\tau)$ and the
superscript $eq$ denotes the corresponding number
in equilibrium. In the quasifree approximation, the absorption of
the $T_{cc}$ can be taken into account by using the
two-body scattering $D(D^*)+\pi\rightarrow D(D^*)+\pi$,
\begin{equation}
\langle \sigma_{T_{cc}\pi\rightarrow DD^*\pi} v_{T_{cc}\pi}\rangle(\tau)=
c_1 \langle \sigma_{D\pi\rightarrow D\pi}
v_{T_{cc}\pi}\rangle(\tau)
+
c_1  \langle \sigma_{D^*\pi\rightarrow D^*\pi}
v_{T_{cc}\pi}\rangle(\tau) \, ,
\end{equation}
where the factor $c_1$ will depend on the configuration of
$T_{cc}$ for which we will consider the following two cases.
\begin{enumerate}
\item {\it Compact configuration}: Compact configuration is
expected when the $T_{cc}$ is composed dominantly of a
color triplet $\bar{q}\bar{q}$ state and a color anti-triplet $cc$
state \cite{potential,Lipkin:1986dw}. Then the decomposition into
two $c \bar{q}$ states will result in the color decomposition
given as \cite{Park:2013fda}
\begin{eqnarray}
T_{cc} =
\frac{1}{\sqrt{3}} \left(D_1D_1^{*}\right)
-\sqrt{\frac{2}{3}}\left(D_8D_8^{*}\right)\, ,
\label{color-decom}
\end{eqnarray}
where $D_1, D_8$ respectively denote the singlet and octet
components of the $c \bar{q}$ color state. Hence, due to the
coupling to color singlet states, we will take $c_1=\frac{1}{3}$.
\item {\it Molecular configuration}: If the diquark correlation is
not strong enough, the $T_{cc}$ could be a molecular
configuration of $D,D^*$ coming from the long range pion exchange
\cite{Manohar:1992nd,Hyodo:2017hue}. For this case we take
$c_1=1$.
\end{enumerate}
The production term of Eq. (\ref{evolve}) has three bodies in the
initial state, and we have approximated it using the
equilibrium condition as derived in Appendix~\ref{appendix-rate}.

\begin{figure}
\includegraphics[width=0.45\textwidth]{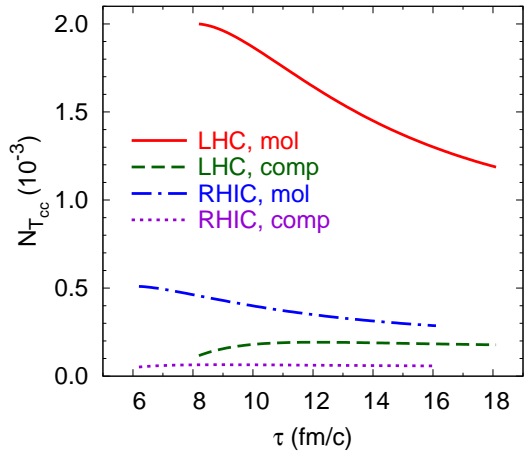}
\caption{The expected time evolution of the $T_{cc}$ abundance in
Pb+Pb collisions at $\sqrt{s_{NN}}=2.76$ TeV at LHC and Au+Au
collisions at $\sqrt{s_{NN}}=200$ GeV at RHIC. } \label{result}
\end{figure}

To obtain the abundance of the $T_{cc}$, we have solved the
rate equation Eq. (\ref{evolve}) with the initial yields
$N_{T_{cc}}(\tau_H)$ given in Table \ref{NTcc}. By using the
equilibrium distributions for $N_D(\tau)$ and $N_{D^*}(\tau)$, the
numerical results are shown in Fig. \ref{result}. Here we have 
used the time dependencies obtained by ideal hydrodynamic
calculations.  Those obtained using viscous hydrodynamics give almost the
same result. In the first term of Eq. (\ref{evolve}), the
absorption rate of the $T_{cc}$ is approximately $0.06$ c/fm
because $\langle\sigma_{D\pi\rightarrow D\pi}v\rangle(\tau)\sim 6$ mb in Fig. \ref{cross} (b) and $n_\pi(\tau)\sim 0.1\,
\mbox{fm}^{-3}$. This alone would lead  to about 45\%
reduction of the abundance as the typical lifetime of the
hadronic phase is 10 fm/c. On the other hand, the production rate
is approximately $\mathcal{O}(10^{-4})$ smaller than the
absorption rate, which can be seen easily from the factor
$N_{T_{cc}}^{eq}(\tau)/[N_D^{eq}(\tau)N_{D^*}^{eq}(\tau)]$.
Hence, its contribution becomes important only at high
density when the numbers of $D,D^*$ mesons are large. Effectively,
the production  depends on the relative abundance between
$N_{T_{cc}}(\tau)$ and $N_{T_{cc}}^{eq}(\tau)$. For molecular
configurations, while
$N_{T_{cc}}(\tau_H)=N_{T_{cc}}^{eq}(\tau_H)$, the equilibrium
number decreases and hence the number of $T_{cc}$ decreases (less
than $42\%$) with time. For a compact multiquark state with
relatively small initial yields, the number of the $T_{cc}$
increases but it remains to be an order of magnitude smaller than
a molecular configuration as the cross section for production
is as small as that for the absorption.

The final yield of the $T_{cc}$ depends strongly on the
initial number at hadronization. Because of the large initial
yield, the expected abundance of the $T_{cc}$ at LHC is
larger than that at RHIC. These results mean that the
numbers of charm quarks and the $T_{cc}$ produced from the
quark-gluon plasma phase are important to determine the final
abundance of the $T_{cc}$. We can conclude that for both
the RHIC and LHC experiments, the large difference between the
statistical and coalescence expectations, obtained assuming that
the $T_{cc}$ is a compact multiquark or molecular
configuration, remains until the kinetic freeze-out.

We have also considered the case that $D$ and $D^*$ are not in
chemical equilibrium. This is important as the total number of
charm quarks is expected to be conserved during the hadronic
phase. The processes where the numbers of $D,D^*$ change are
related to Eq.~(\ref{evolve}), where the absorption of the
$T_{cc}$ is related to the production of $D,D^*$ and its inverse
relation. However, instead of solving the coupled rate equations involving
charmed hadrons, we will consider two extreme cases.
\begin{enumerate}
\item After the chemical freeze-out, the
numbers of $D$ and $D^*$ will be assumed to be constant. \\
\item We will assume that the inelastic cross sections involving
light hadrons are large so that the ratios of charmed hadrons
follow the equilibrium ones until the kinetic freeze-out point.
While extreme, such a scenario seems to be consistent with the
experimental findings for the $K^*/K$ ratios from heavy ion
collisions \cite{Cho:2015qca}. This scenario is easily implemented
by allowing the fugacity $\gamma_c(\tau)$ to depend on time during
the hadronic phase using the following condition:
\begin{eqnarray}
\sum_{D_i=D,D^*,D_s,D_s^*}N_{D_i}(\tau) \, &=& \gamma_c(\tau) \bigg[ N^{0}_D(\tau)+N^{0}_{D^*}(\tau)
+N^0_{D_s}(\tau)+N^{0}_{D_s^*}(\tau)    \bigg] \, , \nonumber\\
&=& \mbox{total number of charm quarks} \, ,
\label{charm-fugacity}
\end{eqnarray}
where $N_{D_i}^{0}(\tau)$'s are the equilibrium numbers given in
Eq.~(\ref{stateq}) without the  fugacity. Once $\gamma_c(\tau)$ is
obtained, one can assume that the individual numbers also satisfy
the similar relations at each time.
\begin{eqnarray}
N_D(\tau) & = & \gamma_c(\tau)  \, N^{0}_D(\tau) \, ,  \nonumber \\
N_{D^*}(\tau) & = & \gamma_c(\tau) \, N^{0}_{D^*}(\tau) \, .
\label{limitingsolution}
\end{eqnarray}
This will guarantee that the charm-anticharm annihilation
processes are small such that the total numbers of charmed and
anticharmed mesons remain constant throughout the hadronic phase.
\end{enumerate}
The correct numbers would be somewhere between the two extreme cases.

\begin{figure}
\includegraphics[width=0.45\textwidth]{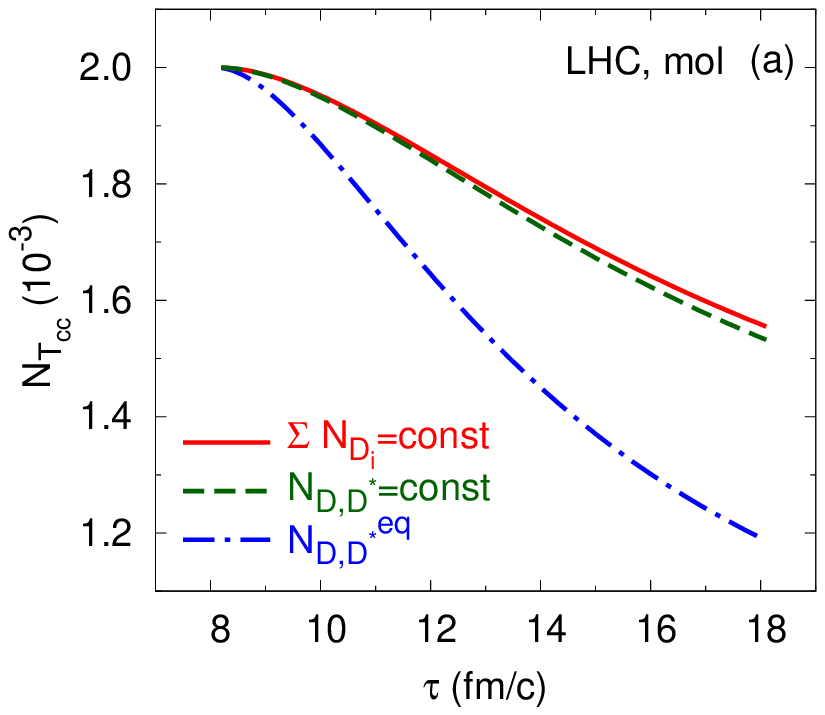}
\includegraphics[width=0.45\textwidth]{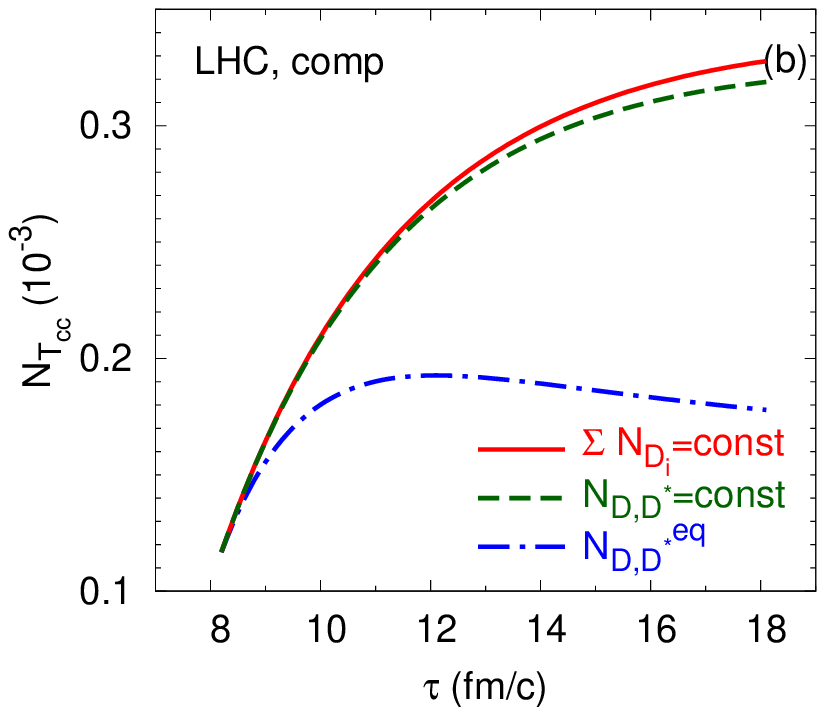}
\caption{$N_{D,D^*}(\tau)$ dependence on the $T_{cc}$ yields for
(a) molecular and (b) compact configurations. } \label{result_mc}
\end{figure}

Fig. \ref{result_mc} shows the results for the two cases. When $D$
and $D^*$ are not in equilibrium, the $T_{cc}$ is more
likely to be produced for both molecular and compact
configurations. In fact, the number of the $T_{cc}$ is
largest in the limit of Eq. (\ref{charm-fugacity}). However,
we still find that even in this extreme limit, the abundance for a
compact multiquark state at the end of the hadronic phase remains
to be a factor of 5 smaller than that for a  molecular
configuration.

\section{Final states}
\label{finalstates}

Here we will list the possible final states that could be
measured to reconstruct the $T_{cc}$ from heavy ion collisions.
The model calculations at present vary on the exact value of the
binding energy. Therefore, we will probe all possibilities
\cite{Lee:2007tn}. It should be noted that one could also look at
the charge conjugate final states and search for
$T_{\bar{c}\bar{c}}$ mesons.
\begin{enumerate}
\item $m_{T_{cc}} \ge m_D+m_{D^*}$: In this case,
\begin{equation}
\label{case1}
T_{cc}~~ \rightarrow  ~~{\rm (a)\,}D^0 +D^{*+}  ~~~{\rm or}~~~
{\rm (b)\,}D^+ +D^{*0} ~~~{\rm or}~~~ {\rm (c)\,}D^++D^++\pi^-.
\end{equation}
As $D^{*+} \rightarrow D^0+\pi^+$ and $D^0 \rightarrow
K^-+\pi^+$, 
(a) can be reconstructed with vertex detectors. 
$D^{*0}$ in (b) may not be easy to detect directly. 
\\
\item $m_D+m_{D^*} \ge m_{T_{cc}} \ge m_D+m_{D} +m_\pi $: This
would be the most likely case for a compact multiquark state.
Then, the virtual $D^{*+}$ component can decay into $D^0 +\pi^+$ so
that a detectable final state would be
\begin{equation}
\label{case2}
T_{cc} ~~\rightarrow~~  D^0 +D^0 +\pi^+ . 
\end{equation}
The final state involving 
$T_{cc} \rightarrow D^0 +D^+ +\pi^0$
would be harder to identify. 
We note that the final state of Eq. (\ref{case2}) is not distinguishable with 
that of Eq. (\ref{case1}) (a).  
\\
\item $ m_{T_{cc}} \le m_D+m_{D} +m_\pi $: In this case, the
virtual $D^*$ component should also decay into $D +\pi$ so that
a detectable final state would be
\begin{equation}
T_{cc} ~~\rightarrow~~  D^0 +K^-+\pi^+ +\pi^+ 
~~~{\rm or}~~~  D^+ +K^-+\pi^++\pi^+ +\pi^- .
\end{equation}
\end{enumerate}
Among all the above cases, Eqs. (\ref{case1}) (c) $(D^++D^++\pi^-)$ and 
(\ref{case2}) $( D^0 +D^0 +\pi^+)$  seem to be the most probable case to reconstruct 
the $T_{cc}$.

\section{Summary}
\label{summary}

We have investigated the hadronic effects on the
$cc\bar{q}\bar{q}$ tetraquark state by focusing on the
$T_{cc}$ multiplicity during the hadronic phase at RHIC and LHC.
In particular, we have considered the absorption by pions and the
inverse process within the quasifree approximation, where the
$T_{cc}$ is considered as a $D,D^*$ state with appropriate
coupling strength depending on whether it has a compact multiquark
or molecular structure. We have extracted the time dependence of
the volume and temperature for the hadronic phase for both the
RHIC and LHC from the hydrodynamic calculations based on the
lattice equation of state with or without viscosity. By solving
the rate equation for the $T_{cc}$ and estimating the changes for
the $D$ and $D^*$ number, we have calculated how much the
structure dependent initial number changes during the hadronic
phase. 
Furthermore,  we have also considered all the possible final states that could be
measured to reconstruct the $T_{cc}$ from heavy ion collisions.  Among all the cases, we find $D^++D^++\pi^-$ and 
$D^0 +D^0 +\pi^+$  to be the most probable case to reconstruct 
the $T_{cc}$. 

For a molecular configuration, where the initial number of 
the $T_{cc}$ is expected to follow the statistical
model prediction, the absorption effect is larger than production and 
reduces the abundance by about
$42\%$. When a compact tetraquark structure is assumed,
the initial number estimated from a coalescence model 
 is an order of magnitude smaller than that from
the statistical model estimate, and hence production 
is larger. However, we find that due to the small
cross section of about 5 mb, the rate of change is not large
enough so that the initial order of magnitude difference in the
assumed abundance is maintained at the end
of the hadronic phase. 
This suggests that measuring the
$T_{cc}$ from heavy ion collisions could also tell us about the
nature of its structure, which could either be a compact multiquark state or a loosely bound molecular configuration.

\section*{Acknowledgements}

This work was supported by the Korea National
Research Foundation under the grant number 2016R1D1A1B03930089, 
by the National Research Foundation of Korea
(NRF) grant funded by the Korea government (MSIP) (No. 2016R1C1B1016270), and 
by the National Research Foundation of Korea
(NRF) grant funded by the Korea government (MSIT) 
(No. 2018R1C1B6008119).

\appendix

\section{2 $\rightarrow$  3 scattering}
\label{appendix-23}

Consider a process where two particles of momenta $q+k_1$ scatter
into 3 particles of momenta $p_1+p_2+k_2$. 
The cross section is written as
\begin{eqnarray}
\sigma_{diss} &= & \frac{1}{2 E_q 2E_{k_1}v_{qk_1} g_qg_{k_1} }
\int \frac{d^3\p_2}{(2\pi)^3 2 E_{p_2}}\frac{d^3\p_1}{(2\pi)^3 2 E_{p_1}}
\frac{d^3\k_2}{(2\pi)^3 2 E_{k_2}} \nonumber\\
&& \times (2 \pi)^4
\delta^4(p_1+p_2+k_2-q-k_1) |\mathcal{M} |^2 \, , \nonumber \\
&= &  \int \frac{d^3\p_2}{(2\pi)^3 2 E_{p_2}}\int d^4k \delta^4(k+p_2-q)
\frac{1}{2E_q2E_{k_1}v_{qk_1} } (2E_k2E_{k_1}v_{kk_1} ) \nonumber \\
&& \times  \frac{1}{2E_k2E_{k_1}v_{kk_1} g_qg_{k_1}  }
\int \frac{d^3\p_1}{(2\pi)^3 2 E_{p_1}}  \frac{d^3\k_2}{(2\pi)^3 2 E_{k_2}}
 (2 \pi)^4 \delta^4(p_1+k_2-k-k_1) |\mathcal{M} |^2
\label{cross-23} \, .
\end{eqnarray}
The matrix element is defined from
\begin{equation}
\mathcal{M} = \langle q,k_1 |p_2, p_1, k_2 \rangle \, ,
\end{equation}
where each state is normalized as
\begin{equation}
\langle q |p \rangle  =  (2 \pi)^3 2E_q \delta^3(\q-\p) \, .
\end{equation}

The quasifree (QF) part is given as
\begin{equation}
\mathcal{M}_{QF} = \langle k,k_1 |p_1, k_2 \rangle \, .
\end{equation}
Quasifree approximation means that all the particles involved are
on-shell.  Therefore, we can approximate in the $q$-rest frame
($\q=0$)
\begin{eqnarray}
|\mathcal{M}|^2 & = & \frac{g_{q}}{g_{k}}
\bigg|\frac{\langle q |p_2, k \rangle}{ \langle k | k \rangle}\bigg|^2
|\mathcal{M}_{QF}|^2 \, , \nonumber \\
& = & \frac{g_{q}}{g_{k}}(2 \pi)^3 2 E_q
\delta^3(\p_2+\k) |\mathcal{M}_{QF}|^2 \, .
\label{mqf}
\end{eqnarray}
If we allow small off-shell effects, as explicitly shown in
Ref.~\cite{quasifree}, one can approximate the right hand side of
Eq.~(\ref{mqf}) as follows:
\begin{eqnarray}
|\mathcal{M}|^2 =\frac{g_{q}}{g_{k}}(2
\pi)^3 2 E_q |\psi(p)|^2 |\mathcal{M}_{QF}|^2 \, ,
\label{mqf2}
\end{eqnarray}
where $\psi(p)$ is the relative wave function of the bound state 
with $p\approx|\p_2|\approx|\k|$.

Substituting Eq.~(\ref{mqf}) into Eq.~(\ref{cross-23}), we obtain
\begin{eqnarray}
\sigma_{diss} &= & \int d^4k \delta^4(k+p_2-q) \frac{E_q}{E_{p_2}}
\frac{1}{2E_q2E_{k_1}v_{qk_1}} (2E_k2E_{k_1}v_{kk_1} ) \nonumber \\
&& \times  \frac{1}{2E_k2E_{k_1}v_{kk_1}g_kg_{k_1} } \int
\frac{d^3\p_1}{(2\pi)^3 2 E_{p_1}}  \frac{d^3\k_2}{(2\pi)^3 2 E_{k_2}} (2
\pi)^4 \delta^4(p_1+k_2-k-k_1) |\mathcal{M}_{QF} |^2 \, . \qquad
\label{cross-23b}
\end{eqnarray}
We assume that $q$ and $p_2$ are at rest and the resonance is
barely bound so that $m_q/2=m_k=m_{p_2}$. Then, we have
\begin{eqnarray}
\sigma_{diss} &= &  \frac{1}{2E_k2E_{k_1}v_{kk_1}g_kg_{k_1} }
\int \frac{d^3\p_1}{(2\pi)^3 2 E_{p_1}}  \frac{d^3\k_2}{(2\pi)^3 2 E_{k_2}}
(2 \pi)^4 \delta^4(p_1+k_2-k-k_1) |\mathcal{M}_{QF} |^2 \, , \nonumber \\
& =& \sigma_{QF} \, .
\end{eqnarray}
Therefore, any addition of thermal factors related to external
particles could be obtained by multiplying the corresponding
thermal factors $f(\p)$.

So far, we have assumed that the QF scattering occurs with only
one constituent. If the interaction occurs with other particles
independently, one can just sum the matrix element. However,
quantum mechanical effects with a specific form of the interaction
is important when interference terms are to be taken care
consistently. The interaction between particle 1,2 and a third
particle 3 with a respective flavor matrix $\lambda_i$, can be
written in general as
\begin{eqnarray}
\lambda_1 \lambda_3+ \lambda_2 \lambda_3 & = & \frac{1}{2}
\bigg[ (\lambda_1+\lambda_2+\lambda_3 )^2-(\lambda_1+\lambda_2)^2-
\lambda_3^2 \bigg] \, , \nonumber \\
& =& 0 \qquad {\rm if~} (\lambda_1+\lambda_2)=0 \, .
\end{eqnarray}
That is, the cross section would be zero if there is no additional
momentum difference in the vertex. A nonzero contribution arises
when there is a derivative acting on the momentum difference
between particle 1 and 2. In this case, the interaction will pick
up a term proportional to the dipole of the system
\cite{quasifree}
\begin{eqnarray}
\lambda_1 \lambda_3 \partial \psi(p) \, .
\end{eqnarray}
After the momentum integral, the matrix element would be of order
$\mathcal{O}(1)$ as the typical momentum the derivative picks up
would be inversely proportional to the size of the wave function.
Therefore, the QF approximation would be good as an order of
magnitude estimate of the cross section even when the total
isospin of the bound state is zero.

\section{Rate equation}
\label{appendix-rate}

In $2\rightarrow 2$ case ($A+B \rightarrow C+D$), the interaction
rate is given by
\begin{eqnarray}
\frac{dN}{Vd\tau}(A+B\rightarrow C+D)&=&g_Ag_B\int \frac{d^3\p_A}{(2\pi)^3}
\frac{d^3\p_B}{(2\pi)^3}f_A(\p_A)f_B(\p_B)v_{AB}\sigma_{A+B\rightarrow C+D}
\, , \nonumber\\
&=&\int \frac{d^3\p_A}{(2\pi)^3 2E_A}\frac{d^3\p_B}{(2\pi)^3 2E_B}
\frac{d^3\p_C}{(2\pi)^3 2E_C}\frac{d^3\p_D}{(2\pi)^3 2E_D} f_A(\p_A)f_B(\p_B)
\nonumber\\
&&\times(2\pi)^4\delta^{(4)}(p_A+p_B-p_C-p_D)
|\mathcal{M}_{A+B\rightarrow C+D}|^2\, .
\end{eqnarray}
Generalizing the second line of the above equation to $M\rightarrow N$ case, 
\begin{multline}
\frac{dN}{Vd\tau}(A_1+A_2+\cdots+A_M\rightarrow
B_1+B_2+\cdots+B_N)= \int \prod_{i=1}^M\frac{d^3\p_{A_i}}{(2\pi)^3
2E_{Ai}}f_{A_i}(\p_{A_i})
\prod_{j=1}^N\frac{d^3\p_{B_j}}{(2\pi)^3 2E_{Bj}}  \\
\times(2\pi)^4\delta^{(4)}(p_{A1}+\cdots+p_{A_M}-p_{B1}-\cdots-p_{B_N})
|\mathcal{M}_{A_1+\cdots+A_M\rightarrow B_1+\cdots+B_N}|^2 \, . 
\label{general}
\end{multline}
Applying Eq. (\ref{general}) to our study,
$T_{cc}+\pi\rightarrow D+D^*+\pi$ and $D+D^*+\pi\rightarrow T_{cc}+\pi$, 
\begin{multline}
\frac{dN}{Vd\tau}(T_{cc}+\pi \rightarrow D+D^*+\pi)=\int \frac{d^3\p_{D}}
{(2\pi)^3 2E_D}\frac{d^3\p_{D^*}}{(2\pi)^32E_{D^*}}\frac{d^3\p_{\pi^f}}
{(2\pi)^3 2E_{\pi^f}}\frac{d^3\p_{T_{cc}}}{(2\pi)^3 2E_{T_{cc}}}
\frac{d^3\p_{\pi^i}}{(2\pi)^3 2E_{\pi^i}} \\
\times f(\p_{T_{cc}})f(\p_{\pi^i}) (2\pi)^4\delta^{(4)}(p_D+p_{D^*}+
p_{\pi^f}-p_{T_{cc}}-p_{\pi^i})|\mathcal{M}_{T_{cc}+\pi
\rightarrow D+D^*+\pi}|^2 \, ,
\end{multline}
and
\begin{multline}
\frac{dN}{Vd\tau}(D+D^*+\pi\rightarrow T_{cc}+\pi)=\int
\frac{d^3\p_{D}} {(2\pi)^3
2E_D}\frac{d^3\p_{D^*}}{(2\pi)^32E_{D^*}}\frac{d^3\p_{\pi^f}}
{(2\pi)^3 2E_{\pi^f}}\frac{d^3\p_{T_{cc}}}{(2\pi)^3 2E_{T_{cc}}}
\frac{d^3\p_{\pi^i}}{(2\pi)^3 2E_{\pi^i}} \\
\times f(\p_D)f(\p_{D^*})f(\p_{\pi^i})
(2\pi)^4\delta^{(4)}(p_D+p_{D^*}+ p_{\pi^i}-p_{T_{cc}}-p_{\pi^f})
|\mathcal{M}_{D+D^*+\pi\rightarrow T_{cc}+\pi}|^2 \, , 
\end{multline}
where $\pi^i$ and $\pi^f$ are incoming and outgoing pions.

Since the transition amplitude for $D+D^*+\pi\rightarrow
T_{cc}+\pi$ is same as that for $T_{cc}+\pi \rightarrow
D+D^*+\pi$, the change of the number of $T_{cc}$ is given by
\begin{multline}
 \frac{dN_{T_{cc}}}{Vd\tau}=\int \frac{d^3\p_{D}}{(2\pi)^3
2E_D}
\frac{d^3\p_{D^*}}{(2\pi)^32E_{D^*}}\frac{d^3\p_{\pi^f}}{(2\pi)^3
2E_{\pi^f}} \frac{d^3\p_{T_{cc}}}{(2\pi)^3
2E_{T_{cc}}}\frac{d^3\p_{\pi^i}}{(2\pi)^3 2E_{\pi^i}} \\
 \times
(2\pi)^4\delta^{(4)}(p_D+p_{D^*}+p_{\pi^f}-p_{T_{cc}}-p_{\pi^i})
|\mathcal{M}_{T_{cc}+\pi \rightarrow D+D^*+\pi}|^2 \\
 \times
[f(\p_D)f(\p_{D^*})f(\p_{\pi^i})-f(\p_{T_{cc}})f(\p_{\pi^f})] \, .
\label{number}
\end{multline}

The scattering cross section for $T_{cc}+\pi \rightarrow
D+D^*+\pi$ is given by
\begin{multline}
\sigma_{T_{cc}+\pi \rightarrow D+D^*+\pi}=\frac{1}{2E_{T_{cc}}2E_{\pi^i}
v_{T_{cc}\pi^i}g_{T_{cc}}g_\pi}\int \frac{d^3\p_{D}}{(2\pi)^3 2E_D}
\frac{d^3\p_{D^*}}{(2\pi)^32E_{D^*}}\frac{d^3\p_{\pi^f}}{(2\pi)^3 2E_{\pi^f}} \\
\times (2\pi)^4\delta^{(4)}(p_D+p_{D^*}+p_{\pi^f}-p_{T_{cc}}-p_{\pi^i})
|\mathcal{M}_{T_{cc}+\pi \rightarrow D+D^*+\pi}|^2 \, .
\end{multline}

\begin{enumerate}

\item {\it Absorption}: We introduce the thermal averaged cross
section defined in Eq. (\ref{thermal-average}) in the text. Then,
the absorption can be written as
\begin{eqnarray}
\frac{dN_{T_{cc}}}{Vd\tau}=- \langle
\sigma_{T_{cc}\pi \rightarrow DD^*\pi} v_{T_{cc}\pi} \rangle
n_{T_{cc}}n_{\pi} \, ,
\end{eqnarray}
where
\begin{eqnarray}
n=\frac{N}{V}=g\int\frac{d^3\p}{(2\pi)^3} f(\p) \, .
\end{eqnarray}

\item {\it Production}: Instead of working out the three body
cross section, using detailed balance, we
will take it to be of the following form:
\begin{eqnarray}
\frac{dN_{T_{cc}}}{Vd\tau}= \langle \sigma_{T_{cc}\pi \rightarrow
DD^*\pi} v_{T_{cc}\pi} \rangle  n^{eq}_{T_{cc}} \, \frac{n_D
n_{D^*}}{n^{eq}_D n^{eq}_{D^*}} \, n_{\pi} \, .
\end{eqnarray}

\end{enumerate}

Collecting the absorption and production terms,
\begin{eqnarray}
\frac{dN_{T_{cc}}}{Vd\tau}= \langle
\sigma_{T_{cc}\pi \rightarrow DD^*\pi} v_{T_{cc}\pi} \rangle n_{\pi}
\bigg( n^{eq}_{T_{cc}}\frac{n_D n_{D^*}}{n^{eq}_D n^{eq}_{D^*}} - n_{T_{cc}}
\bigg) \, .
\end{eqnarray}
This is the rate equation we will be using in Eq.~(\ref{evolve}).

\end{document}